\documentclass[aps,prb,superscriptaddress,twocolumn,groupedaddress, showpacs]{revtex4}

\usepackage{graphicx}
\usepackage{amsmath}
\usepackage{amsfonts}
\usepackage{amssymb}
\usepackage{units}
\usepackage{color}

\begin{document}

\title{Epitaxially stabilized iron thin films via effective strain relief from steps}

\author{T. Miyamachi}
\email[]{toshio.miyamachi@issp.u-tokyo.ac.jp}
\affiliation{The Institute for Solid State Physics, The University of Tokyo, Kashiwa, Chiba 277-8581, Japan}

\author{S. Nakashima}
\affiliation{The Institute for Solid State Physics, The University of Tokyo, Kashiwa, Chiba 277-8581, Japan}

\author{S. Kim}
\affiliation{The Institute for Solid State Physics, The University of Tokyo, Kashiwa, Chiba 277-8581, Japan}

\author{N. Kawamura}
\affiliation{The Institute for Solid State Physics, The University of Tokyo, Kashiwa, Chiba 277-8581, Japan}
\affiliation{NHK Science $\&$ Technology Research Laboratories (STRL), Kinuta, Tokyo 157-8510, Japan}

\author{Y. Tatetsu}
\affiliation{Department of Physics, The University of Tokyo, Hongo, Tokyo 113-0033, Japan}

\author{Y. Gohda}
\affiliation{Department of Physics, The University of Tokyo, Hongo, Tokyo 113-0033, Japan}
\affiliation{Department of Materials Science and Engineering, Tokyo Institute of Technology, Yokohama 226-8502, Japan}

\author{S. Tsuneyuki}
\affiliation{The Institute for Solid State Physics, The University of Tokyo, Kashiwa, Chiba 277-8581, Japan}
\affiliation{Department of Physics, The University of Tokyo, Hongo, Tokyo 113-0033, Japan}

\author{F. Komori}
\email[]{komori@issp.u-tokyo.ac.jp}
\affiliation{The Institute for Solid State Physics, The University of Tokyo, Kashiwa, Chiba 277-8581, Japan}


\begin{abstract}

We show a new way to stabilize epitaxial structures against transforming bulk stable phases for Fe thin films on a vicinal Cu(001) surface. Atomically-resolved observations by scanning tunneling microscopy reveal that high-density Cu steps serve as strain relievers for keeping epitaxially-stabilized Fe fcc(001) lattice even at a transient thickness towards the bulk stable bcc(110) lattice. Spectroscopic measurements further clarify the intrinsic electronic properties of the fcc Fe thin film in real space, implying electronic differences between 6 and 7 monolayer thick films induced by the modification of the lattice constant in the topmost layers.

\end{abstract}

\date{\today}
\maketitle

\section{INTRODUCTION}
\label{sec:INTRODUCTION}
 Ferromagnetic thin films have been studied both for fundamental interest in their magnetism depending on the size and shape, and for their high potential for the device applications. Considerable efforts have been made so far to understand structure-dependent electronic and magnetic properties of the thin films with various crystal structures. In the case of epitaxially-grown thin films, a close interplay between the film and substrate is highlighted at the interface. On one hand, due to the designed matching of the lattice constant, ideal electronic and magnetic properties can be sustained in the thin film  \cite{Himpsel_PRL_1991, Yuasa_APL_2006, Kawagoe_APEX_2009}. On the other hand, the existing lattice mismatch can enhance the magnetic moments \cite{Thiele_PRB_1996, Lee_PRB_2010}, magnetic anisotropy \cite{Carcia_APL_1985, Krams_PRL_1992, Chen_PRL_1992}, and spin polarization \cite{Maruyama_PRB_1992}. In both cases, an appropriate choice of the substrate is crucial to stabilize the epitaxial thin films with desired functionalities. However, transformation of the crystal structure from the epitaxial one to the bulk stable one with increasing coverage complicates the evaluation of intrinsic electronic and magnetic properties of the epitaxially-stabilized lattice.

\begin{figure}[!b]
\includegraphics[width=8.0cm]{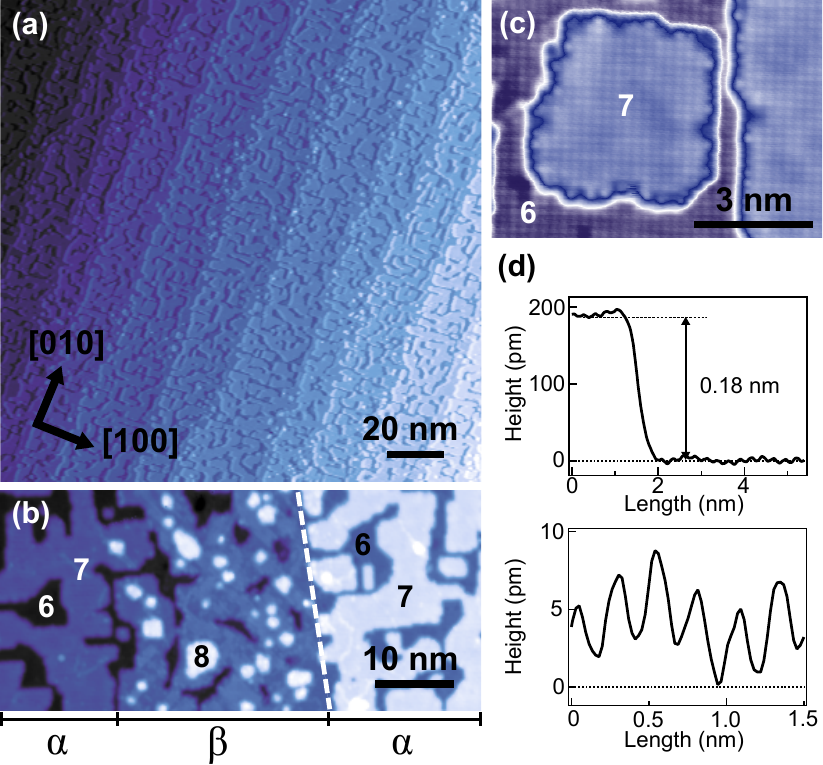} 
\caption{(Color online) (a) STM image of 6.9 ML Fe thin film on Cu(001) with high-density steps along [010] direction. (b) STM image with the regions $\alpha$ and $\beta$. The white dotted line represents the Cu step. The upper terrace on the right side of the step is the region $\alpha$. The left side of the step is the lower terrace consisting of the region $\beta$ (center) and region $\alpha$ (left). (c) Atomically-resolved STM image in the region $\alpha$. (d) Typical STM line profiles of the region $\alpha$ showing 1 ML height (upper panel) and atomic periodicity (lower panel) of the Fe fcc(001) lattice.}
\label{Fig1}
\end{figure}

\begin{figure*}[!]
\includegraphics[width=14.5cm]{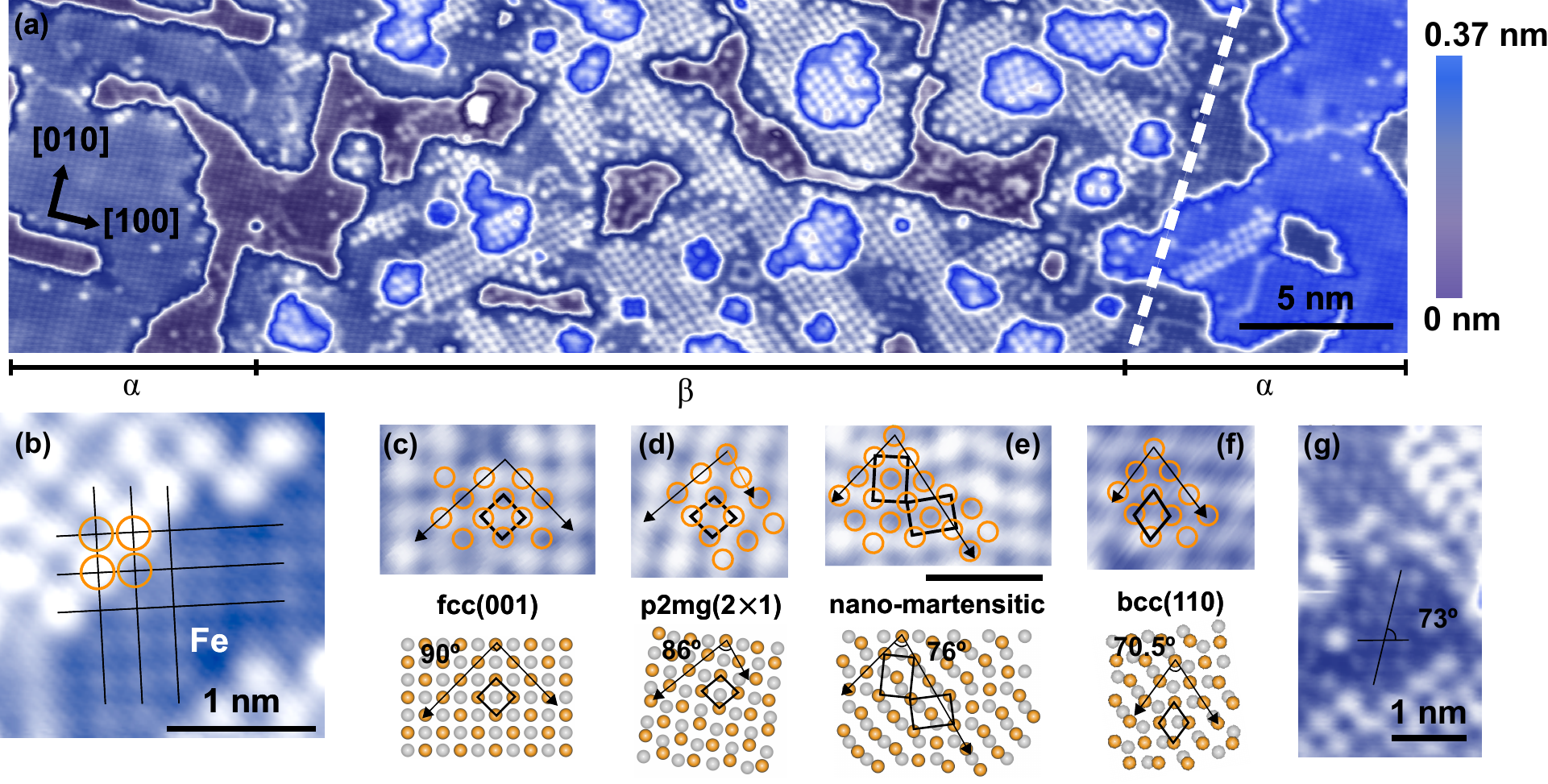} 
\caption{(Color online) (a) Atomically-resolved STM image in the regions $\alpha$ and $\beta$. The white dotted line represents the Cu step. The left of the step edge is the lower terrace. (b) Atomically-resolved STM image in the region $\beta$ with the Fe fcc(001) lattice and adsorbates. The Fe fcc(001) lattice (grids and circles) is superimposed on the STM image. (c)-(f) Surface reconstructions with adsorbates in the region $\beta$. (upper panels) STM images with the superimposition of underlying Fe lattices (circles). The scale bar corresponds to 1 nm. (lower panels) Corresponding schematic models of the underlying Fe lattices. The orange and gray circles represent 1st and 2nd Fe layers, respectively. (g) Atomically-resolved STM image in the region $\beta$ with the Fe bcc(110) lattice.}
\label{Fig2}
\end{figure*}

 We here report on an additional mechanism to further stabilize the epitaxial film by step edges that serve as strain relievers. High-density steps on the substrate can prevent the epitaxial lattice from transforming towards the bulk stable phase with increasing coverage. We chose Fe thin films grown epitaxially on a vicnal Cu(001) surface, whose structural properties have been rarely investigated so far while the growth on a flat Cu(001) are well defined, as a prototypic system to demonstrate the concept proposed avobe. The system in the fcc phase has experimentally and theoretically attracted a great interest over two decades because of its complex magnetic structures \cite{Qian_PRL_2001, Asada_PRL_1997, Meyerheim_PRL_2009, Sandratskii_PRB_2010}. It has been proposed that the top layer ferromagnetically couples to the second layer with antiferromagnetically-coupled underneath layers. However, the structural transformation from fcc to bcc Fe implied by the appearance of the surface reconstruction \cite{Fowler_PRB_1996, Biedermann_SurfSci_2004} is still a bottleneck to experimentally achieve a common understanding of the magnetic structure of the fcc Fe thin films.
 
 In the present study, we unambiguously demonstrate that the fcc lattice of epitaxial Fe thin films on Cu(001) is stabilized by high-density steps acting as strain relievers. Atomically-resolved observations with scanning tunneling microscopy (STM) provide the information inaccessible by spatially-averaged techniques that the Fe surface close to the step edge preserves the fcc(001) lattice, whereas the surface far from the step edge shows the reconstructed structures transforming towards the bulk stable bcc(110) lattice. Furthermore, combining atomic-layer resolved scanning tunneling spectroscopy (STS), we reveal that the observed electronic differences between 6 and 7 monolayer (ML) thick fcc Fe thin films are derived from the lattice compression and expansion of the topmost layers.

\section{EXPERIMENTAL DETAILS}
\label{sec:DETAILS}
 All the measurements were performed with a low-temperature STM in an ultra-high vacuum ($<$ 2.0 $\times$ 10$^{-11}$ Torr) at 80 K using electrochemically-etched W tips. A clean Cu(001) surface was prepared by several cycles of Ar$^{+}$ sputtering and subsequent annealing at 770 K. The crystal was miscut by 0.4$^{\circ}$ towards the [100] direction, which results in the average terrace width of about 20 nm on the vicinal Cu(001) surface with a broad distribution of the terrace width between 8 and 140 nm. Iron was deposited on the clean Cu(001) at room temperature by molecular beam epitaxy (MBE) from a high-purity Fe rod (99.998$\%$) with the growth rate of 0.8 ML/min. The base pressure of the MBE chamber was below 9.0 $\times$ 10$^{-11}$ Torr during the Fe deposition. The first derivative spectra of the tunneling current, dI/dV, were recorded using a lock-in technique with a bias-voltage modulation of 20 mV and 512 Hz. 
 

\section{RESULTS AND DISCUSSION}
\label{sec: RESULTS}
 Figure 1(a) displays a large scale STM image of a vicinal Cu(001) surface after 6.9 ML Fe deposition on average.  The average width of the terraces is 20 nm in this region. The Fe thin film on the vicinal substrate grows in a layer-by-layer mode as on a flat Cu(001) substrate \cite{Giergiel_PRB_1995}. Most of the surface is covered by 6 or 7 ML with the surface fractions of 15 and 80 $\%$, respectively. Hereafter we call this the region $\alpha$. In a magnified STM image near the step edge as shown in Fig. 1(b), high-density 8 ML islands are noticeable on the lower terraces, in contrast to the region $\alpha$ on the upper terrace with a negligible amount of the 8 ML islands. We call the lower terrace with the 8 ML islands near the step edge the region $\beta$. Figure 1(c) displays an atomically-resolved STM image of the region $\alpha$, where square (1$\times$1) lattices are clearly visible \cite{footnote1}. The STM line profiles of the region $\alpha$ shown in Fig. 1(d) give the monolayer height and the atomic periodicity of about 0.18 and 0.26 nm, respectively. These values are consistent with the well-known Fe fcc(001) lattice on Cu(001) \cite{Giergiel_PRB_1995}. Interestingly, we observed only the fcc(001) lattice in the region $\alpha$, in contrast to 7 ML Fe thin films on a flat Cu(001) substrate, where several types of the surface reconstruction are confirmed \cite{Biedermann_SurfSci_2004}. Thus, the region $\alpha$ with the stabilized fcc(001) lattice is intrinsic to the Fe thin film on the vicinal substrate.

 An atomically-resolved STM image of the lower Fe terrace near the step edge  on a 80 nm wide terrace, including both the regions $\alpha$ and $\beta$, is shown in Fig. 2(a)  \cite{footnote1}. Ordered areas with adsorbates are noticeable on the surface. The density of the adsorbates in the region $\beta$ is much higher than that in the region $\alpha$, which indicates the large difference in the sticking coefficient between the regions $\alpha$ and $\beta$.
 
 The sites of adsorbates can be deduced by superimposing underlying substrate lattices \cite{Miyamachi_e-JSSN_2011}. Figure 2(b) displays a part of the region $\beta$, where the adsorbates forming a c(2$\times$2) structure and the Fe fcc(001) surface with the square (1$\times$1) lattice are visible. The image clearly indicates that the adsorbates locate on top of the underlying Fe atoms. Considering the species of the residual gas in the vacuum chamber, the adsorbates could be carbon monoxide since it adsorbs on top of Fe atoms \cite{Abe_PRB_2008}. In addition to the c(2$\times$2) structure, the adsorbates forming three types of reconstructed structures were observed as shown in upper panels of Figs. 2(c-f). In all the cases, the adsorbates regularly cover half of the surface, which in turn enables us to determine the lattice structures of the underlying Fe surface in the region $\beta$. Here, we assume that the adsorbates are on top of the underlying Fe atoms and form a c(2$\times$2) structure on each Fe lattices.

 Lower panels of Figs. 2(c-f) display the models of the underlying Fe surface extracted from the observed reconstructed structures. We find that, in addition to the fcc(001) lattice, the Fe surface in the region $\beta$ consists of the areas with p2mg(2$\times$1), nano-martensitic \cite{Biedermann_Appl_Phys_A_2004}, and bcc(110) structures. The fractions of the fcc(001) and reconstructed p2mg(2$\times$1) structures are higher than the nano-martensitic structure. The bcc(110) structure covers very narrow areas (less than a few $\%$). Considering bond angles between fcc(001) and bcc(110), the existence of the p2mg(2$\times$1) and nano-martensitic structures could be in the process of a structural transition from fcc to bcc Fe. These reconstructed structures appearing only in the region $\beta$ were previously reported for Fe thin films on a flat Cu(001) substrate with a similar thickness \cite{Biedermann_SurfSci_2004, Biedermann_Appl_Phys_A_2004}, which indicates that they grow in the similar manner. Note that these reconstructed surfaces are not stabilized by the adsorbates. We can clearly see the bare Fe bcc(110) lattice (bond angle: $\sim$ 73$^{\circ}$) in the region $\beta$ as shown in Fig. 2(g). These facts are in contrast to the adsorbate-induced structural transition from fcc to bcc Fe islands on Cu(111) \cite{Biedermann_PRB_2006}. The slight deviation of the angel evaluated in the experiment from the ideal one of 70.5$^{\circ}$ could be caused by the lattice relaxation at the surface as previously reported from the ion beam triangulation (71.5$^{\circ}$) \cite{Bernhard_PRL_2005} and LEED (72.1$^{\circ}$) \cite{Wuttig_SurfSci_1993} studies, and slightly by the thermal drift of the STM image.

\begin{figure}[!t]
\includegraphics[width=8.5cm]{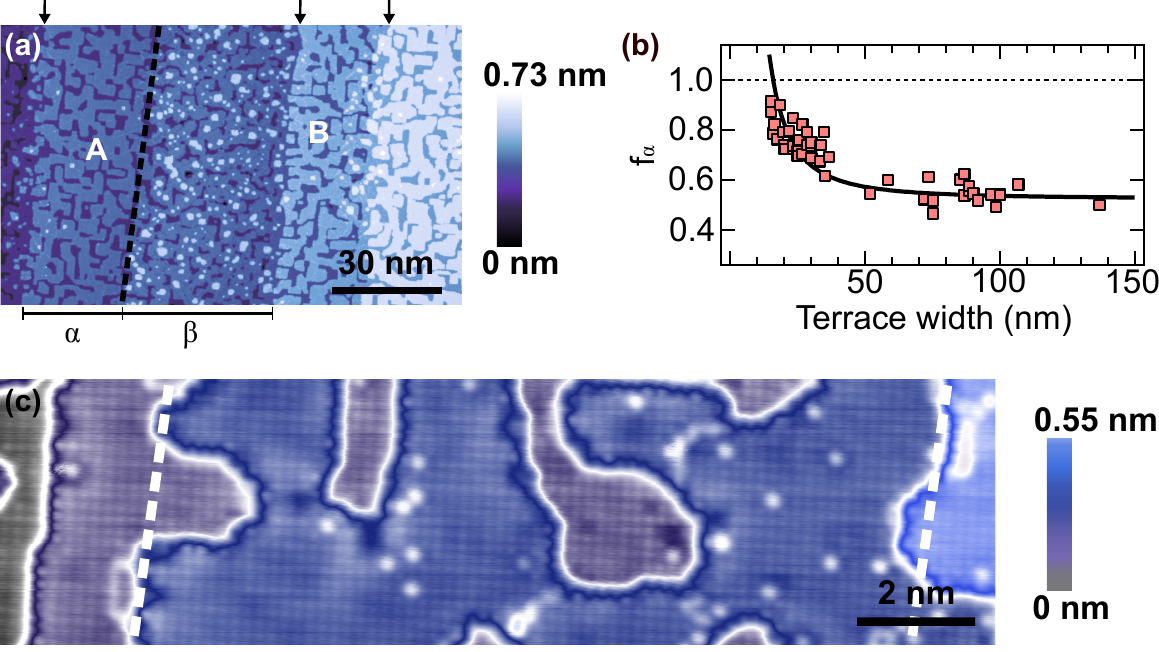} 
\caption{(Color online) (a) STM image with wide (A) and narrow (B) Fe terraces.  The dashed line represents the boundary between the regions $\alpha$ and $\beta$ on the wide terrace A. The arrows show the step positions. (b) Statistical plot of the fraction of the region $\alpha$, $f_\alpha$, as a function of the terrace width. The solid line is a fitting with an inverse square law. (c) Atomically-resolved STM image of a narrow Fe terrace. The Fe fcc(001) lattice is stabilized in the whole terrace. The white dashed lines represent the Cu steps.}
\label{Fig3}
\end{figure}

 Figure 3(a) displays two adjacent terraces with different widths. We find that the fraction of the region $\alpha$, $f_\alpha$, strongly depends on the terrace width. The terrace A is 70 nm wide and $f_\alpha$ is $\sim$ 40 $\%$, whereas on the 30-nm-wide terrace B, $f_\alpha$ is more than $\sim$ 80 $\%$ on average. On the latter terrace, the region $\beta$ is seen only in the vicinity of the step to the upper terrace. We also notice that the boundary between the regions $\alpha$ and $\beta$ is parallel to the step direction as indicated by the black dashed line in Fig. 3(a). This fact unambiguously indicates that the width of the region $\alpha$ in a fixed width terrace is determined by the distance from the step edges. To understand the origin determining the fraction of the region $\alpha$, $f_\alpha$ is plotted as a function of the terrace width as shown in Fig. 3(b). It increases as the terrace width decreases. According to the continuum elasticity theory on the propagation of the strain produced by a step edge \cite{Pimpinelli_1998}, the strain decays as $1/r^2$ at a long distance r from the edge. The observed tendency following the inverse square law of the terrace width for $f_\alpha$ should be related to the elastic propagation of the strain from the step edges. The fitting using an inverse square law gives the value of $f_\alpha$ $\sim$ 1.0 at the terrace width of 15 nm. Indeed, the region $\beta$ is absent on the terrace narrower than 10 nm [see Fig. 3(c)].
 
 We attribute the stabilization of the epitaxial Fe fcc(001) lattice on the upper terrace near the step edges to the lattice expansion of the substrate. Generally, the step edge plays a crucial role as a strain reliever, and the fcc lattice with a larger atomic spacing than the bcc lattice can be stabilized near the step edges on the upper terrace. In the region far from the step edge on a wide terrace, the lattice expansion near the step edges is less effective to stabilize the fcc lattice of the Fe thin film. This could result in the emergence of the local p2mg(2$\times$1), nano-martensitic and bcc(110) lattices, i.e., the formation of the region $\beta$ on a wide terrace as in the case of the Fe films on a flat substrate \cite{Biedermann_SurfSci_2004}. In the region $\beta$, the formation of 8 ML small islands could be promoted due to the inhomogeneous growth of the four different crystal structures and the increased lattice strain in the Fe thin film. Note that, in the region $\beta$, the Fe fct (bct) lattice can exist, which has a comparable lateral lattice constant but expanded interlayer spacing of the order of 10 pm with respect to Fe fcc (bcc) lattice \cite{Muller_PRL_1995}. These height differences could increase the lattice strain in the region $\beta$. However, we find no such height differences in the region $\beta$ with atomically resolved STM observations. 

 The effective strain relief of the fcc lattice from the step edges allows us to study the intrinsic electronic properties of the fcc Fe in real space with STS. We performed STS near the Fermi energy to discuss electronic properties of the regions $\alpha$. The upper panel of Fig. 4(a) displays dI/dV spectra recorded on 6 and 7 ML in the region $\alpha$. The 7 ML spectrum exhibits peaks at + 0.15 V and $-$ 0.45 V, respectively. The energy positions of these peaks well correspond to Fe 3d minority states of the fcc Fe thin films on Cu(001) \cite{Montano_PRL_1987, Montano_PRB_1988}. The 6 ML spectrum shows the similar electronic structures, but with the slightly narrower peak separation (+ 0.09 V and $-$ 0.35 V) and weaker peak intensity at the positive bias compared with the 7 ML spectrum. The electronic difference can be homogeneously seen in the dI/dV map at + 0.2 V as shown in Fig. 4(b). Note that the 6 ML areas are minor in the present film. At a glance, the narrower peak separation and weaker peak intensity of the 6 ML spectrum seems to be strange since the comparable magnetic moments and thus similar electronic structures are expected for the top layer \cite{Sandratskii_PRB_2010}.

\begin{figure}[!t]
\includegraphics[width=8.0cm]{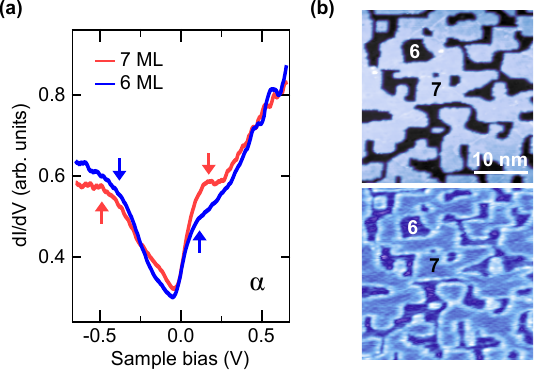} 
\caption{(Color online) (a) 6 and 7 ML dI/dV spectra recorded in the region $\alpha$. (b) (upper panel) STM image and  (lower panel) corresponding dI/dV map recorded at + 0.2 V in the region $\alpha$.}
\label{Fig4}
\end{figure}

\begin{figure}[!t]
\includegraphics[width=8.5cm]{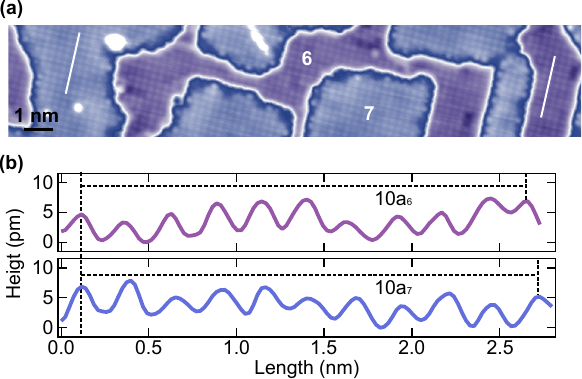} 
\caption{(Color online) (a) Atomically-resolved STM image in the region $\alpha$. (b) STM line profiles of the atomic corrugation of (upper panel) 6 and (lower panel) 7 ML along the white lines in (a).}
\label{Fig5}
\end{figure}

To explore the origin of the observed difference in the electronic structures between 6 and 7 ML in the region $\alpha$, we refocus on the stabilized Fe fcc(001) lattice. The magnetic moment of the ferromagnetic fcc Fe thin film is known to become larger with increasing the lattice constant, which is caused by the enhanced localization of 3d states \cite{Moruzzi_PRB_1989}. Thus, the even slight change of the lattice constant can also modify the electronic structures of the fcc Fe thin films.
 
 Firstly, the lateral lattice constants of 6 and 7 ML in the region $\alpha$, $a_{6}$ and $a_{7}$, respectively, are evaluated from the line profiles in one atomically-resolved STM image as shown by the white lines in Fig. 5(a) to minimize the influence of the STM thermal drift. Figure 5(b) displays atomic corrugations equivalent to 10 fcc(001) lattices on 6 and 7 ML. We find that 10 $a_{7}$ (2.61 nm) is slightly longer than 10 $a_{6}$ (2.55 nm), which gives $a_{7}$ and $a_{6}$ of 0.261 and 0.255 nm on average. The observed lateral lattice expansion of 7 ML compared to 6 ML by $\sim$ 2.4 $\%$ could be attributed to the presence of the steps to the 6 ML regions, which themselves are compressed as in the case of a MnN nanostructure on Cu(001) \cite{Liu_PRL_2007}. The modified lattice constant of 7 ML relative to that of 6 ML could lead to their different electronic structures as shown in Fig. 4 (a).
 
 

\section{CONCLUSION}
\label{sec: CONCLUSION}
We show that utilizing the strain relief from substrate step edges by controlling the step density will successfully stabilize the epitaxial fcc(001) lattice of Fe thin films on Cu(001). Especially on the narrow Cu terraces ($<$ $\sim$10 nm), the Fe fcc(001) lattice is stabilized over the whole terrace, while the mixture of reconstructed structures appears at the regions far from the step edges on the relatively-wide Cu terraces. On a structurally controlled surface, the intrinsic electronic structures of the fcc Fe are experimentally investigated in real space. We reveal the slight difference in the electronic structures between fcc Fe 6 and 7 ML. In combination with atomically-resolved observations, the electronic differences could be interpreted in terms of the lattice expansion of 7 ML relative to 6 ML. Future works in conjunction with theoretical calculations taking atomic-scale features on the surface into consideration will further clarify the intrinsic magnetic properties of fcc Fe thin films.

\section{ACKNOWLEDGEMENT}
\label{sec: ACKNOWLEDGEMENT}
This work is partly supported by JSPS KAKENHI for Young Scientists (A), Grant No. 16H05963, for Scientific Research (B), Grant No. 26287061, the Hoso Bunka Foundation, Shimadzu Science Foundation, Iketani Science and Technology Foundation and the Elements Strategy Initiative Center for Magnetic Materials (ESICMM) under the outsourcing project of MEXT.

\bibliographystyle{prl}
\bibliography{references}

\begin{references}
\bibitem[{foo({\natexlab{1}})}]{footnote1}
\bibinfo{note}{Note that the color contrast of atomically-resolved STM images is emphasized such that the surface lattice structures in each layer and adsorbates are clearly visible.}
\end{references}


\end{document}